\pdfoutput=1
\documentclass[twoside,11pt]{article}

\usepackage{authblk}
\usepackage{fancyhdr}
\usepackage{hyperref}
\usepackage{algorithm}
\usepackage{amsmath}

\newtheorem{theorem}{Theorem}[section]
\newtheorem{example}{Example}[section]
\newtheorem{remark}{Remark}[section]

\newcommand{\B}[1]{{\bf #1}}
\newcommand{\R}[1]{{\rm #1}}

\begin{document}

\title{
    Computing Sparse Jacobians and Hessians Using Algorithmic Differentiation
}

\author[1]{Bradley M. Bell}
\affil[1]{IHME, University of Washington, Seattle, USA, bradbell@seanet.com}

\author[2]{Kasper Kristensen}
\affil[2]{DTU Aqua, Technical University of Denmark, DK, kaskr@dtu.dk}

\maketitle

\begin{abstract}
Stochastic scientific models and
machine learning optimization estimators have a large number of variables;
hence computing large sparse Jacobians and Hessians is important.
Algorithmic differentiation (AD) greatly reduces the programming effort
required to obtain the sparsity patterns and values for these matrices.
We present forward, reverse, and subgraph methods
for computing sparse Jacobians and Hessians.
Special attention is given the the subgraph method because it is new.
The coloring and compression steps are not necessary
when computing sparse Jacobians and Hessians using subgraphs.
Complexity analysis shows that for some problems the subgraph
method is expected to be much faster.
We compare C++ operator overloading implementations of the methods
in the ADOL-C and CppAD software packages
using some of the MINPACK-2 test problems.
The experiments are set up in a way that makes them easy to run
on different hardware, different systems, different compilers,
other test problem and other AD packages.
The setup time is the time to record the graph,
compute sparsity, coloring, compression, and optimization of the graph.
If the setup is necessary for each evaluation,
the subgraph implementation has similar run times for sparse Jacobians
and faster run times for sparse Hessians.
\end{abstract}

{\small




} 

\section{Introduction}
This paper concerns the computation of sparse Jacobians and Hessians
for a function $f : \B{R}^n \rightarrow \B{R}^m$ that can be evaluated
using a computer program.
We use $x \in \B{R}^n$ to denote the inputs
and $y \in \B{R}^m$ the outputs of the program.

\subsection{The Computational Graph Representation}
\label{SectionComputationalGraph}
We use $v \in \B{R}^\ell$ to denote all intermediate values
that depend on the inputs
\[
v = ( v_1 , \ldots , v_\ell )^\R{T} \in \B{R}^n
\]
The independent variable subvector (input vector) is
\[
x = (v_1 , \ldots , v_n )^\R{T} \in \B{R}^n \; .
\]
The dependent variable subvector (output vector) is
\[
y = ( v_{\ell-m+1}, \ldots , v_\ell )^\R{T} \in \B{R}^m \; .
\]
We decompose the computations to a sequence of elementary functions
$\phi_k : \B{R}^2 \rightarrow \B{R}$.
For $k = n+1$ to $\ell$
\[
v_k = \phi_k ( v_{a[k]} , v_{b[k]} )
\]
where the positive integer $a[k]$ and $b[k]$ are the variable index
of the left and right operands for the $k$-th elementary function.
It follows that $a[k] < k$ and $b[k] < k$, i.e.,
computation of the $k$-th variable
only depends on the independent variables and
variables that have been computed previously.
The variables are nodes in the \textit{computation graph}.
Each elementary operation corresponds to two arcs,
one from variable $v_{a[k]}$ to variable $v_k$
and the other from variable $v_{b[k]}$  to variable $v_k$.
We also write $a[k] \prec k$ and $b[k] \prec k$ which means
variable $k$ directly depends on variables $a[k]$ and $b[k]$.
This is very similar to the notation in \cite[Section 2.2]{GriewankEDP2008}.
Some references call nodes $a[k] , b[k]$ predecessors of node $k$ and
node $k$ a successor of $a[k] , b[k]$; for example see
\cite{Gower2014}, \cite{Wang2016}.

We represent a unary elementary function using
$a[k] = b[k]$ together with a binary function
$\phi_k( v_{a[k]} , v_{b[k]} )$ that does not depend on it second argument.
One example of a unary representation is
$\phi_k( v_{a[k]}, v_{b[k]} ) = \sin( v_{a[k]} )$.
Another example is
$\phi_k ( v_a[k], v_b[k] ) = v_{a[k]}$ which
can be used to set $y_i = v_{\ell-m+i} = v_{a[\ell-m+i]}$.
It is not necessary to have the extra node for $v_{\ell - m + i}$ in the graph
for this purpose but it makes this presentation simpler.

\begin{example}
\label{GraphExample}
Consider the computer program
defined by the following pseudo code:

\begin{tabbing}
\hspace*{4cm}\=\hspace*{1cm}\=\hspace*{2cm}\=\kill
\> \R{function} $y$ = $f(x)$                        \\
\> \> $( v_1 , v_2 , v_3 )$ \> = $x$                \\
\> \>  $v_4$                \> = $v_1 + v_2$        \\
\> \>  $v_5$                \> = $v_3 * v_4$        \\
\> \>  $y$                  \> = $( v_4 , v_5 )$    \\
\end{tabbing}

\noindent
The corresponding function
$f : \B{R}^3 \rightarrow \B{R}^2$ is given by
\[
\begin{array}{rcl}
f_1(x) & = &  x_1 + x_2 \\
f_2(x) & = & x_3 (x_1 + x_2) \\
\end{array}
\]

\noindent
The following is a diagram of the corresponding computational graph:
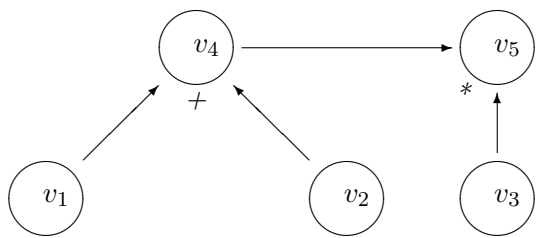
\begin{center}
\setlength{\unitlength}{1cm}
\begin{picture}(8, 4)
\put(1,1){\circle{1}} \put(1,1){ \makebox(0,0)[c]{$v_1$} }
\put(5,1){\circle{1}} \put(5,1){ \makebox(0,0)[c]{$v_2$} }
\put(7,1){\circle{1}} \put(7,1){ \makebox(0,0)[c]{$v_3$} }
\put(3,3){\circle{1}} \put(3,3){ \makebox(0,0)[c]{$v_4$} }
\put(1.5,1.5){\vector(1,1){1}}
\put(4.5,1.5){\vector(-1,1){1}}
\put(3,2.3){ \makebox(0,0)[r]{+} } 
\put(7,3){\circle{1}} \put(7,3){ \makebox(0,0)[c]{$v_5$} }
\put(7,1.6){\vector(0,1){.8}}
\put(3.6,3){\vector(1,0){2.8}}
\put(6.4,1.9){ \makebox(0,)[c]{*} }
\end{picture}
\end{center}
If the last line of the program pseudo code were changed to $y = v_4$,
then $f : \B{R}^3 \rightarrow \B{R}$, $f(x) = x_3 ( x_1 + x_2)$
and the diagram above would not change.

\end{example}

The dimension of the range of $\phi_k$ might be greater than one,
and its domain might be greater than two. For example,
in some operator overloaded AD tools it is possible to record
the operations for one function and make that function an elementary operation
in another function. This is closely related to checkpointing; see
\cite[Tables 12.3, 12.4]{GriewankEDP2008}.
We only include binary elementary functions in order to simplify
this presentation.

\subsection{Dependency Relation}
There may be a difference between sparsity patterns and dependency patterns.
For example, suppose that
\[
\phi_k ( v_{a[k]}, v_{b[k]} ) = \left\{ \begin{array}{ll}
    0 & \mbox{if} \; v_{a[k]} \leq 0
    \\
    1 & \mbox{otherwise}
\end{array} \right.
\; .
\]
It follows that, where $\phi_k$ is differentiable, its derivative is zero.
Some AD packages might return zero for the derivative everywhere while others
might return $( + \infty , 0 )$ when $v_{a[k]} = 0$.
In any event, $\phi_k (v_{a[k]}, v_{b[k]})$ depends on the value of $v_{a[k]}$.
Hence, a dependency pattern
might include more possibly non-zero elements than a Jacobian sparsity pattern.
For the purpose of this presentation, we do not distinguish
Jacobian sparsity patterns and dependency patterns.
We use the notation $a[k] \prec k$ and $b[k] \prec k$ to denote the fact
that $v_k$ directly depends on $v_{a[k]}$ and $v_{b[k]}$.
Furthermore, we use $\prec^*$ to denote
the transitive closure of this relation; e.g.,
if $p \prec q$ and $q \prec k$ then $p \prec^* k$;
see \cite[Section 2.2]{GriewankEDP2008}.
Note that the $\phi_k (v_{a[k]}, v_{b[k]})$
defined above does not depend on $v_{b[k]}$;
i.e., it represents a unary function.
In the text below,
unary functions correspond to the case where $\phi_k$ does not depend
on its second argument and $a[k] = b[k]$.

\section{Sparse Jacobians}
In this section we present a
forward algorithm, reverse algorithm, and reverse subgraph algorithm
that compute Jacobian sparsity patterns.
We also present an algorithm that obtains a sorted subgraph
and discuss computing Jacobians and the computational complexity
of the subgraph algorithms.

\subsection{Forward Mode Sparsity}
We use $J$ to denote the subset of independent variable indices
that are of interest, $J \subset \{ 1 , \ldots , n \}$.
For the independent variable indices,
$k \in \{ 1 , \ldots , n \}$,
we define $X_k$ to be the singleton $\{ k \}$, if $k \in J$,
otherwise the empty set.
For $k > n$ we define $X_k$ to be the indices of independent variables
in $J$ that the variable $v_k$ depends on; i.e.,
\[
X_k = \left\{ \begin{array}{ll}
    \{ k \} & \mbox{if} \; k \leq n \; \mbox{and} \; k \in J
    \\
    \emptyset & \mbox{if} \; k \leq n \; \mbox{and} \; k \notin J
    \\
    \{ j : j \in J \; \mbox{and} \; j \prec^* k \} & \mbox{otherwise}
\end{array} \right.
\; .
\]
The forward Jacobian sparsity calculation
is given by Algorithm~\ref{AlgorithmForwardJacobianSparsity};
see \cite[Eq. (7.4)]{GriewankEDP2008}.
\begin{algorithm}
\label{AlgorithmForwardJacobianSparsity}
\caption{Forward Jacobian Sparsity}
\begin{tabbing}
\hspace{2em}\=\hspace{2em}\=\hspace{4em}\= \kill
\{ $X_1$ , \ldots , $X_\ell$ \} = \B{function}($J$) \\
\> \B{for} $k = 1 , \ldots , n$ \\
\> \> \B{if} $k \in J$ \> \B{set} $X_k = \{ k \}$  \\
\> \> \B{else}         \> \B{set} $X_k = \emptyset$ \\
\> \B{for} $k = n+1 , \ldots , \ell$ \\
\> \> \B{set} $X_k = X_{a[k]} \cup X_{b[k]}$
\end{tabbing}
\end{algorithm}
Note that for unary functions, $a[k] = b[k]$ and the union
in the algorithm is not necessary.
The sparsity pattern for the Jacobian $f^{(1)} (x)$ is given by the
observation that
for $i = 1 , \ldots , m$,
$j \in J$, and
$x \in \B{R}^n$,
\[
    [ f_i^{(1)} (x) ]_j \neq 0 \; \Rightarrow \; j \in X_{\ell-m+i}
    \; .
\]
We note that there are $\ell$ sets $X_k$ and
at most $|J|$ elements in each of these sets
where $|J| \leq n$ is the number of elements in set $J$.

\subsection{Reverse Mode Sparsity}
We use $I$ to denote the subset of dependent variable indices
that are of interest, $I \subset \{ 1 , \ldots , m \}$.
For the corresponding variable indices
$k \in \{ \ell - m + 1 , \ldots , \ell \}$
we define $Y_k$ to be the singleton $\{ k - \ell + m \}$,
if $k - \ell + m \in I$,
otherwise the empty set.
For $k \leq \ell - m$ we define $Y_k$ to be the indices of dependent variables
that the variable in $I$ that are affected by $v_k$; i.e.,
\[
Y_k = \left\{ \begin{array}{ll}
    \{ k - \ell + m \}
    & \mbox{if} \; k > \ell - m \; \mbox{and} \; k - \ell + m \in I
    \\
    \emptyset
    & \mbox{if} \; k > \ell - m \; \mbox{and} \; k - \ell + m \notin I
    \\
    \{ i \in I : \; k \prec^* \ell - m + i \}
    & \mbox{otherwise}
\end{array} \right.
\; .
\]
We use $\stackrel{\cup}{=}$ for the operator that sets the left hand side
to the right hand side union the previous value of the left hand side.
The reverse Jacobian sparsity calculation
is given by Algorithm~\ref{AlgorithmReverseJacobianSparsity};
see \cite[Eq. (7.8)]{GriewankEDP2008}.
\begin{algorithm}
\label{AlgorithmReverseJacobianSparsity}
\caption{Reverse Jacobian Sparsity}
\begin{tabbing}
\hspace{2em}\=\hspace{2em}\=\hspace{8em}\= \kill
\{ $Y_1$ , \ldots , $Y_\ell$ \} = \B{function}($I$) \\
\> \B{for} $k = \ell, \ldots , \ell - m + 1$ \\
\> \> \B{if} $k - \ell + m \in I$ \> \B{set} $Y_k = \{ k - \ell + m \}$ \\
\> \> \B{else}                    \> \B{set} $Y_k = \emptyset$ \\
\> \B{for} $k = \ell - m , \ldots , 1$ \\
\> \> \B{set} $Y_k = \emptyset$ \\
\> \B{for} $k = \ell - m, \ldots , 1$ \\
\> \> \B{set} $Y_{a[k]} \stackrel{\cup}{=} Y_k$ \\
\> \> \B{set} $Y_{b[k]} \stackrel{\cup}{=} Y_k$
\end{tabbing}
\end{algorithm}
Note that for unary functions, $a[k] = b[k]$ and the second union
in  the algorithm is not necessary.
The sparsity pattern for the Jacobian $f^{(1)} (x)$ is given by the
observation that
for $i \in I$,
$j = 1 , \ldots , n$, and
$x \in \B{R}^n$,
\[
    [ f_i^{(1)} (x) ]_j \neq 0 \; \Rightarrow \; i \in Y_j
    \; .
\]
We note that there are $\ell$ sets $Y_k$ and
at most $|I|$ elements in each of these sets
where $|I| \leq m$ is the number of elements in set $I$.

\subsection{Reverse Subgraph Sparsity}
\label{SectionReverseSubgraphSparsity}
We call this a reverse subgraph method because on each pass through
the graph it only visits the nodes that affect a selected dependent variable.
(A forward subgraph method would only visit nodes that are affected by
a selected independent variable.)
We present two versions of the algorithm.
The first is the `pattern only' variant which finds
the sparsity pattern of the Jacobian matrix.
The second variant is an extension that finds
the actual subgraph required to calculate the non-zero values of the Jacobian.

We use $I \subset \{ 1 , \ldots , m\}$
to denote the subset of dependent variable indices that are of interest.
We use $J \subset \{ 1 , \ldots , n\}$
to denote the subset of independent variable indices that are of interest.
The set $S_i$ accumulates the independent variables in $J$ that affect
the dependent variable $y_i = v_{\ell - m + i}$.
The stack $K$ contains the nodes in the subgraph
that have not yet been processed
for this independent variable index $i$.
Let $\{ X_k \}$ be the output corresponding to
Algorithm~\ref{AlgorithmForwardJacobianSparsity} with input $J$.
It is not necessary to have the sequence of sets $\{ X_k \}$,
just the integer sequence $\{ c_k \}$ initialized by
\[
\mbox{
\B{if} $X_k = \emptyset$ \B{set} $c_k = m+1$ \B{else} \B{set}  $c_k = 0$
} \; .
\]
The value $c_k = i < m+1$ is used to indicate that node $k$
has already been processed for this independent variable index $i$.
The reverse subgraph Jacobian sparsity calculation is defined by
Algorithm~\ref{AlgorithmReverseSubgraphJacobianSparsity}.
\begin{algorithm}
\label{AlgorithmReverseSubgraphJacobianSparsity}
\caption{Reverse Subgraph Jacobian: Return Sparsity Pattern}
\begin{tabbing}
\hspace{2em}\=\hspace{2em}\=\hspace{2em}\=\hspace{2em}\= \kill
$\{ S_i : i \in I \}$ = \B{function}$(I, c_1 , \ldots , c_\ell )$ \\
\> \B{for} $i \in I$ \\
\> \> \B{set} $\text{done}=i$, $\text{ignore}=m+1$ \\
\> \> \B{set} $K = \emptyset$ \\
\> \> \B{set} $S_i = \emptyset$ \\
\> \> \B{push} $\ell - m + i$ \B{into} $K$  \\
\> \> \B{while} $K \neq \emptyset$ \\
\> \> \> \B{pop} $k$ \B{from} $K$ \\
\> \> \> \B{if} $c_{a[k]} \notin \{ \text{done} , \text{ignore} \}$ \\
\> \> \> \> \B{set} $c_{a[k]} = \text{done}$ \\
\> \> \> \> \B{if} $a[k] \leq n$ \B{set} $S_i \stackrel{\cup}{=} \{ a[k] \}$ \\
\> \> \> \> \B{else} \B{push} $a[k]$ \B{into} $K$ \\
\> \> \> \B{if} $c_{b[k]} \notin \{ \text{done} , \text{ignore} \}$ \\
\> \> \> \> \B{set} $c_{b[k]} = \text{done}$ \\
\> \> \> \> \B{if} $b[k] \leq n$ \B{set} $S_i \stackrel{\cup}{=} \{ b[k] \}$ \\
\> \> \> \> \B{else} \B{push} $b[k]$ \B{into} $K$
\end{tabbing}
\end{algorithm}

The input value for the sequence $\{ c_k \}$ can be computed using
Algorithm~\ref{AlgorithmForwardJacobianSparsity}
with the sets $X_k$ replaced by integer values $c_k$ that are
$m+1$ for empty $X_k$ and $0$ for non-empty $X_k$.
The complexity of this calculation is $O( \ell )$,
$\ell$ is the number of nodes in the graph.
Note that for unary functions, $a[k] = b[k]$ and the condition for the
second if block in the algorithm is false.
The sparsity pattern for the Jacobian $f^{(1)} (x)$ is given by the
observation that
for $i \in I$, $j \in J$, and $x \in \B{R}^n$,
\[
    [ f_i^{(1)} (x) ]_j \neq 0 \; \Rightarrow \; j \in S_i
    \; .
\]
We note that there are $|I|$ sets $S_i$,
at most $|J|$ elements in each of these sets,
and $|I| \leq m$ , $|J| \leq n$.
We contrast this with the forward and reverse mode sparsity patterns
which have $\ell$ sets; i.e., a set for each node in the graph.

\begin{remark}
\label{RemarkReverseSubgraphJacobianSparsity}
Algorithm \ref{AlgorithmReverseSubgraphJacobianSparsity}
is a non-recursive depth-first search in reverse direction
from each dependent variable.
It uses a mark vector $c$, of already processed nodes,
to avoid multiple placements of the same node in the stack $K$.
The mark vector need not be cleared between consecutive searches
because the dependent variable index is used as unique mark for each search.
This is important when the subgraph is much smaller than the full graph.
The `pop $k$ from $K$' in the algorithm could be replaced by
extracting any element from $K$ and the algorithm would still work.
However, extracting the most recently pushed element (the pop)
reduces the worst case space requirement for $K$ from $l$,
the size of the full graph, to the maximum vertex depth of the graph.
\end{remark}

\subsection{Reverse Subgraph Sorting}
\label{SectionReverseSubgraphSorting}
Algorithm~\ref{AlgorithmReverseSubgraphJacobianSparsity}
computes the sparsity patterns $\{ S_i : i \in I \}$.
For each $i$,
the subgraph corresponding to the node order of the pops from $K$
would have to be sorted in dependency order to be used for computations.
For example, to calculate reverse mode derivatives along the subgraph
a dependency order is required.
In a worst case scenario, the subgraphs are as big as the full graph and
the cost of a reverse sweep is $O(l)$.
The cost of sorting the subgraph is $O(l \times log(l))$.
Hence, the sorting could asymptotically become expensive
compared to other parts of the algorithm.
It is therefore relevant to avoid the sort.
With a minor modification of
Algorithm~\ref{AlgorithmReverseSubgraphJacobianSparsity}
we can directly obtain a dependency sorted subgraph.
If $v_{d[1]}, \ldots , v_{d[|G|]}$ is the ordering for a subgraph $G$,
we say that $G$ is \textit{dependency sorted} or
\textit{topologically sorted} if
\[
    v_{d[p]} \prec^* v_{d[q]} \Rightarrow p < q \; .
\]
The main change to the algorithm is as follows:
When the top of the stack is a node,
for which all the nodes it depends on are already in the subgraph,
the top node of the stack is moved to the end of the subgraph.
A new mark $-i$ is used to signify that a node has been moved to the subgraph.
The result is that the nodes in the subgraph are always dependency sorted.

\begin{algorithm}
\label{AlgorithmReverseSubgraphJacobianSubgraph}
\caption{Reverse Subgraph Jacobian: Return Sorted Subgraph}
\begin{tabbing}
\hspace{2em}\=\hspace{2em}\=\hspace{2em}\=\hspace{2em}\=\hspace{2em}\=\hspace{2em}\=\hspace{2em}\= \kill
$ \{ G_i : i \in I \}$ = \B{function}$(I, c_1 , \ldots , c_\ell )$ \\
\> \B{for} $i \in I$ \\
\> \> \B{set} $\text{visited}=i$, $\text{done}=-i$, $\text{ignore}=m+1$ \\
\> \> \B{set} $G_i = \emptyset$, $K = \emptyset$ \\
\> \> \B{push} $\ell - m + i$ \B{into} $K$  \\
\> \> \B{while} $K \neq \emptyset$ \\
\> \> \> \B{set} $k=$ \B{top} (K) \\
\> \> \> \B{comment} If all inputs to node $k$ are done move $k$ to $G_i$ \\
\> \> \> \B{if} $c_{a[k]} \in \{\text{done}, \text{ignore}\}$  \B{and} $c_{b[k]} \in \{\text{done}, \text{ignore}\}$ \\
\> \> \> \> \B{pop} $k$ \B{from} $K$ \\
\> \> \> \> \B{push} $k$ \B{into} $G_i$ \\
\> \> \> \> \B{set} $c_{k} = \text{done}$ \\
\> \> \> \B{else} \\
\> \> \> \> \B{comment} $v_{\max}$ may depend on $v_{\min}$ \\
\> \> \> \> \B{for} $\nu = \max(a[k],b[k]), \min(a[k],b[k])$ \\
\> \> \> \> \> \B{if} $c_{\nu} \notin \{ \text{visited}, \text{done}, \text{ignore} \}$ \\
\> \> \> \> \> \> \B{if} $\nu \leq n$ \\
\> \> \> \> \> \> \> \B{push} $\nu$ \B{into} $G_i$ \\
\> \> \> \> \> \> \> \B{set} $c_{\nu}=\text{done}$ \\
\> \> \> \> \> \> \B{else} \\
\> \> \> \> \> \> \> \B{push} $\nu$ \B{into} $K$ \\
\> \> \> \> \> \> \> \B{set} $c_{\nu} = \text{visited}$ \\
\end{tabbing}
\end{algorithm}

\begin{remark}
\label{RemarkReverseSubgraphJacobianSubgraph}
Algorithm \ref{AlgorithmReverseSubgraphJacobianSubgraph}
can be used to find all the subgraphs, one by one, as needed
to perform a full Jacobian calculation.
When a $G_i$ has been used for a sweep
it is not needed anymore and can be discarded.
Note that $G_i$ is only guarantied to be dependency sorted.
In particular, one cannot assume that the independent variable indices
(node indices $\leq n$)
in the sparsity pattern of the $i$-th row of the Jacobian are
at the beginning of $G_i$.
\end{remark}

\subsection{Computing Jacobians}
\label{SectionComputingJacobians}
Once a sparsity pattern for a Jacobian is available,
a row (column) compression technique could be used to
by reverse mode (forward mode) to multiple rows (columns)
of the Jacobian during one pass of the computational graph.
This requires an approximate solution of a graph coloring problem; see
\cite[Eq. 8.6 and Section 8.3]{GriewankEDP2008},
\cite{Coleman1988}.

The reverse subgraph Jacobian method does not require
coloring or a compression step.
In this way it is similar to the edge pushing algorithm; see
\cite{Petra2018}, \cite{Gower2014}.
Reference \cite{Gower2014} defines the apex-induced subgraph
corresponding to a fixed node as the nodes that the fixed node depends on
(plus the corresponding arcs).
The subgraphs in this paper are apex-induced subgraphs corresponding to
dependent variables.
The subgraph Jacobian calculation method for a selected dependent variable
is the same as for normal reverse pass,
except that only the subset of nodes
that affect a selected dependent variable $y_i$ are processed.
The subgraph and normal reverse mode calculations are so similar
that {\small CppAD} uses iterators to implement both calculations
with the same code; see \cite{BellCppAD}.

\subsection{Complexity of the Subgraph Methods}
\label{SectionComplexitySubgraphMethods}

The asymptotic complexity of the subgraph algorithms
\ref{AlgorithmReverseSubgraphJacobianSparsity} and
\ref{AlgorithmReverseSubgraphJacobianSubgraph} is easy to assess.
Both algorithms loop across subgraphs and nodes in each subgraph.
For each subgraph, each subgraph node is visited once in
Algorithm~\ref{AlgorithmReverseSubgraphJacobianSparsity} and
twice in Algorithm~\ref{AlgorithmReverseSubgraphJacobianSubgraph}.
The work per node is bounded by a constant
independent of the computational graph
because there are no more than two arguments (incoming edges)
and one result (variable) for each node.
It follows that the complexity of both algorithms is
\[
    O \left( \sum_{i \in I} |G_i| \right)  + O( \ell ) \; ,
\]
whether computing patterns only or numerical entries of the Jacobian.
The set $I$ is the independent variables of interest
and $|I|$ is less than or equal $m$.
The term $O( \ell )$ is for the complexity of initializing the sequence
$c_1$ , \ldots , $c_\ell$.
If few independent variables are of interest,
the term $ \sum_{i \in I} | G_i |$ could be less than $\ell$,
the number of nodes in the entire graph.

The formula above tells us that, for a given problem,
the efficiency of the subgraph method is not directly determined by
sparsity of the Jacobian matrix.
What really matters is the amount of overlap between the subgraphs.
The less overlap (smaller $|G_i \cap G_j|$) the faster are the algorithms.
This is in contrast with the graph coloring approach for which
the efficiency is determined by the sparsity pattern of the Jacobian.
We highlight these differences by two simple examples:

\begin{example}
\label{ExampleSubgraphEfficient}
The Jacobian $f^{(1)} (x)$ for this example is a dense matrix.
Let $A$ be a random $n$-by-$n$ matrix and consider
the matrix vector multiplication function $f(x) = A x$.
The computational graph essentially consists of
$n^2$ multiply-add instructions.
The size of the full graph is thus $O(n^2)$.
The size of the $i$-th subgraph is $O(n)$
because $y_i$ is only affected by row $i$ of $A$.
The time to calculate the Jacobian by each of the methods
using $n$ reverse sweeps is:

\begin{center}
\begin{tabular}{ll}
\B{Method}      & \B{Bound} \\
$n$ full sweeps & $O(n^3)$  \\
Coloring        & $O(n^3)$  \\
Algorithm~\ref{AlgorithmReverseSubgraphJacobianSparsity} & $O(n^2 \log(n))$ \\
Algorithm~\ref{AlgorithmReverseSubgraphJacobianSubgraph} & $O(n^2)$
\end{tabular}
\end{center}
Clearly, no method can be faster than $O(n^2)$ for this problem.
Although this example was chosen to show the benefit of the subgraph methods,
it demonstrates that there exist non-sparse problems where
Algorithm~\ref{AlgorithmReverseSubgraphJacobianSubgraph}
is asymptotically optimal.
\end{example}

\begin{example}
\label{ExampleSubgraphNotEfficient}
The Jacobian $f^{(1)} (x)$ for this example is the identity matrix
plus a matrix that is zero except for column $n$.
Suppose that for $k = 1 , \ldots , n$,
$\phi_k$ is a unary function (in the sense of this paper) and
\[
    v_{n + k} = \phi_k ( v_{n+k-1} , v_{n+k-1} ) \; .
\]
Recall that $v_n = x_n$ and
define $f: \B{R}^n \rightarrow \B{R}^n$ by
\[
    y_k = v_{2n} + x_k \; ,
\]
for $k = 1 , \ldots, n$.
The computational graph of $f$ and all its subgraphs are of size $O(n)$.
The sparsity pattern of the Jacobian is
$\{n\}$ for the $n$-th row and
$\{n,k\}$ for rows $k = 1 , \ldots , n-1$.
This Jacobian can be recovered using a combination of
one forward pass and one reverse pass.
The asymptotic complexity of the different methods is:

\begin{center}
\begin{tabular}{ll}
\B{Method}      & \B{Bound} \\
$n$ full sweeps & $O(n^2)$  \\
Coloring        & $O(n)$    \\
Algorithm~\ref{AlgorithmReverseSubgraphJacobianSparsity} & $O(n^2 \log(n))$ \\
Algorithm~\ref{AlgorithmReverseSubgraphJacobianSubgraph} & $O(n^2)$
\end{tabular}
\end{center}
The subgraph method is inefficient for this problem
because of the high subgraph overlap.
The coloring approach is most efficient for this example
because of the special sparsity pattern.
\end{example}

The examples \ref{ExampleSubgraphEfficient} and
\ref{ExampleSubgraphNotEfficient} are also useful for
testing that implementations of a subgraph algorithm scales
as expected for various values of $n$.

\section{Sparse Hessians}
In this section we consider computing the sparsity pattern for the
Hessian of
\begin{equation}
\label{EquationGofX}
    g(x) = \sum_{i=1}^m w_i f_i (x) \; ,
\end{equation}
where $w \in \B{R}^m$.
We use $I$ to denote the
dependent variable indices for which $w_i \neq 0$; i.e.,
\begin{equation}
\label{EquationIofW}
i \in I
\Leftrightarrow
i \in \{ 1 , \ldots , m \} \; \mbox{and} \; w_i \neq 0 \; .
\end{equation}
In addition, we use $J$ to denote the subset of independent variables
that are of interest,
$J \subset \{ 1 , \ldots , n \}$.
Let $\{ X_k \}$ be the output corresponding to
Algorithm~\ref{AlgorithmForwardJacobianSparsity} with input $J$.
Let $\{ Y_k \}$ be the output corresponding to
Algorithm~\ref{AlgorithmReverseJacobianSparsity} with input $I$.
It is not necessary to have the sequence of sets $\{ Y_k \}$,
just the reverse mode activity analysis
Boolean sequence $\{ d_k \}$ defined by
\[
d_k = ( Y_k \neq \emptyset ) \;  .
\]

Here and below $\partial_p h(u)$ is the partial of
$h$ with respect to the $p$-th component of its argument vector
evaluated at $u$.
We use $\partial_{p,q}$ to abbreviate $\partial_p \partial_q$.
We say that the elementary function $\phi_k$ is
\textit{left nonlinear} if $\partial_{1,1} \phi_k (u)$ is possibly
non-zero for some $u$.
It is
\textit{right nonlinear} if $\partial_{2,2} \phi_k (u)$ is possibly
non-zero for some $u$.
It is
\textit{jointly nonlinear} if $\partial_{1,2} \phi_k (u)$ is possibly
non-zero for some $u$.
We assume that $\partial_{1,2} \phi_k (u) = \partial_{2,1} \phi_k (u)$.

\subsection{Forward Mode Sparsity}
Forward mode for function, derivative, and Hessian values
starts with the zero, first, and second
order values for the independent variables;
i.e., $x_j$, $\dot{x}_j$, $\ddot{x}_j$ for $j = 1, \ldots , n$.
It computes the zero, first, and second order values for the
other variables using the following equations
for $k = n+1 , \ldots , \ell$:
\begin{equation}
\label{EquationForwardHessian}
\begin{aligned}
v_k
& = \phi_k ( v_{a[k]} , v_{b[k]} )
\\
\dot{v}_k
& =
\partial_1 \phi_k ( v_{a[k]} , v_{b[k]} ) \dot{v}_{a[k]}
+
\partial_2 \phi_k ( v_{a[k]} , v_{b[k]} ) \dot{v}_{b[k]}
\\
\ddot{v}_k
& =
\partial_1 \phi_k ( v_{a[k]} , v_{b[k]} ) \ddot{v}_{a[k]}
+
\partial_2 \phi_k ( v_{a[k]} , v_{b[k]} ) \ddot{v}_{b[k]}
\\
& +
\partial_{1,1} \phi_k ( v_{a[k]} , v_{b[k]} ) \dot{v}_{a[k]}^2
+
\partial_{2,2} \phi_k ( v_{a[k]} , v_{b[k]} ) \dot{v}_{b[k]}^2
\\
& +
2 \;
\partial_{1,2} \phi_k ( v_{a[k]} , v_{b[k]} ) \dot{v}_{a[k]} \dot{v}_{b[k]}
\; .
\end{aligned}
\end{equation}

The forward mode Hessian sparsity calculation is defined by
Algorithm~\ref{AlgorithmForwardHessianSparsity}.
This is similar to the algorithm \cite[Algorithm II]{Walther2008}.
One difference is using $\{ d_k \}$
to avoid the `dead end' nodes mentioned in the reference and nodes
that are not included in the Hessian because the corresponding $w_i$ is zero;
see Eq~\ref{EquationGofX} and Eq~\ref{EquationIofW}.
This is probably the reason that the \texttt{adolc} implementation has a larger
\textit{nnz} (more possibly non-zero values) than the other implementations in
Table~\ref{TableDeptfgNoSetup} and Table~\ref{TableDeptfgYesSetup}.

The set $N_j$, in the algorithm, accumulates the nonlinear interactions
between the $j$-th independent variable and other independent variables.
\begin{algorithm}
\label{AlgorithmForwardHessianSparsity}
\caption{Forward Hessian Sparsity}
\begin{tabbing}
\hspace{2em}\=\hspace{2em}\=\hspace{2em}\=\hspace{2em}\= \kill
\{ $N_1$ , \ldots , $N_n$ \} =
    \B{function}$(X_1, d_1 , \ldots , X_\ell, d_\ell )$ \\
\> \B{for} $j = 1, \ldots , n$ \B{set} $N_j = \emptyset$ \\
\> \B{for} $k = n+1, \ldots , \ell$ \B{if} $d_k$ \\
\> \> \B{if} $\phi_k$ is left nonlinear \\
\> \> \> \B{for} $j \in X_{a[k]}$ \B{set} $N_j \stackrel{\cup}{=} X_{a[k]}$ \\
\> \> \B{if} $\phi_k$ is right nonlinear \\
\> \> \> \B{for} $j \in X_{b[k]}$ \B{set} $N_j \stackrel{\cup}{=} X_{b[k]}$ \\
\> \> \B{if} $\phi_k$ is jointly nonlinear \\
\> \> \> \B{for} $j \in X_{a[k]}$ \B{set} $N_j \stackrel{\cup}{=} X_{b[k]}$ \\
\> \> \> \B{for} $j \in X_{b[k]}$ \B{set} $N_j \stackrel{\cup}{=} X_{a[k]}$ \\
\end{tabbing}
\end{algorithm}
The nonlinear interactions are initialized as empty.
This corresponds to the second order values
for the independent variables being zero; i.e., $0 = \ddot{x} \in \B{R}^\ell$.
In the case where $\phi_k$ is jointly nonlinear and left nonlinear,
the algorithm \cite[Algorithm II]{Walther2008} uses the fact that
$X_k = X_{a[k]} \cup X_{b[k]}$ to combine two of the unions into one.
A similar optimization is done for the case where
$\phi_k$ is jointly nonlinear and right nonlinear,

The sparsity pattern for the Hessian $g^{(2)} (x)$ is given by the
observation that
for $j \in J$, $p \in J$, and
$x \in \B{R}^n$,
\[
    [ g^{(2)} (x) ]_{j,p} \neq 0 \; \Rightarrow \; p \in N_j
    \; .
\]
Given the second order forward mode equation for
$\ddot{v}_k$ in Eq~\ref{EquationForwardHessian},
a proof for this assertion would be similar to the proof
for Algorithm~\ref{AlgorithmReverseHessianSparsity}.
The Boolean vector $d$ has length $\ell$.
There are $\ell$ sets $X_k$ and
at most $|J|$ elements in each of these sets.
There are $n$ sets $N_j$ and
at most $|J|$ elements in each of these sets.

\subsection{Reverse Mode Sparsity}
The reverse mode Hessian sparsity calculation is defined by
Algorithm~\ref{AlgorithmReverseHessianSparsity}.
This is similar to the table \cite[Table 7.4]{GriewankEDP2008},
but includes more general nonlinear binary functions; e.g.,
$\R{pow}(x, y) = x^y$.
In addition, the algorithm and proof show how to extend the algorithm
to functions with more than two arguments.

\begin{algorithm}
\label{AlgorithmReverseHessianSparsity}
\caption{Reverse Hessian Sparsity}
\begin{tabbing}
\hspace{2em}\=\hspace{2em}\=\hspace{2em}\=\hspace{2em}\= \kill
\{ $M_1$ , \ldots , $M_\ell$ \} =
    \B{function}$(X_1, d_1 , \ldots , X_\ell, d_\ell )$ \\
\> \B{for} $k = 1, \ldots , \ell$ \B{set} $M_k = \emptyset$ \\
\> \B{for} $k = \ell, \ldots , n+1$ \B{if} $d_k$ \\
\> \> \B{set} $M_{a[k]} \stackrel{\cup}{=} M_k$ \\
\> \> \B{set} $M_{b[k]} \stackrel{\cup}{=} M_k$ \\
\> \> \B{if} $\phi_k$ is left nonlinear \\
\> \> \> \B{set} $M_{a[k]} \stackrel{\cup}{=} X_{a[k]}$ \\
\> \> \> \B{set} $M_{b[k]} \stackrel{\cup}{=} X_{a[k]}$ \\
\> \> \B{if} $\phi_k$ is right nonlinear \\
\> \> \> \B{set} $M_{a[k]} \stackrel{\cup}{=} X_{b[k]}$ \\
\> \> \> \B{set} $M_{b[k]} \stackrel{\cup}{=} X_{b[k]}$ \\
\> \> \B{if} $\phi_k$ is jointly nonlinear \\
\> \> \> \B{set} $M_{a[k]} \stackrel{\cup}{=} X_{b[k]}$ \\
\> \> \> \B{set} $M_{b[k]} \stackrel{\cup}{=} X_{a[k]}$ \\
\end{tabbing}
\end{algorithm}
As with Algorithm~\ref{AlgorithmForwardHessianSparsity}
when $\phi_k$ is both left nonlinear and jointly nonlinear
(or right nonlinear and jointly nonlinear)
two of the unions in Algorithm~\ref{AlgorithmReverseHessianSparsity}
can be combined into one.
We include a theorem and proof for this algorithm below.

\begin{theorem}
\label{TheoremReverseHessianSparsity}
For $j \in J$, $p \in J$, and $x \in \B{R}^n$,
\[
    [ g^{(2)} (x) ]_{j,p} \neq 0 \; \Rightarrow \; p \in M_j
    \; .
\]
\end{theorem}

\noindent
\textit{Proof}:
We define the sequence of scalar valued functions
$F_\ell$,  \ldots , $F_n$ by
\[
    F_\ell( v_1 , \ldots , v_\ell )
    =
    \sum_{i=1}^m w_i v_{\ell - m + i}
    \; ,
\]
and for $k = \ell, \ldots , n+1$,
\[
    F_{k-1} ( v_1 , \ldots , v_{k-1} )
    =
    F_k [ v_1 , \ldots , v_{k-1} , \phi_k ( v_{a[k]} , v_{b[k]} ) ]
    \; .
\]
The function $F_n (x)$ is the same as $g(x)$.
Reverse mode computes the derivatives $F_k$
with respect to its arguments for $k = \ell - 1 , \ldots , n$.
The derivative of $F_n$ with respect to its arguments
is equal to $g^{(1)} (x)$ and is the final value for
$\bar{x} = ( \bar{v}_1 , \ldots , \bar{v}_n )$ in the algorithm below.
We use $\stackrel{+}{=}$ for the operator that sets the left hand side
to the right hand side plus the previous value of the left hand side.
\begin{tabbing}
\hspace{2em}\=\hspace{2em}\=\hspace{2em}\=\hspace{4em}\= \kill
\> \B{for} $k = 1 , \ldots , \ell-m$ \B{set} $\bar{v}_k = 0$ \\
\> \B{for} $i = 1 , \ldots , m$ \B{set} $\bar{v}_{\ell - m + i} = w_i$ \\
\> \B{for} $k = \ell , \ldots , n+1$ \\
\> \> \B{set} \(
\bar{v}_{a[k]} \stackrel{+}{=}
    \partial_1 \phi_k ( v_{a[k]} , v_{b[k]} ) \bar{v}_k
\) \\
\> \> \B{set} \(
\bar{v}_{b[k]} \stackrel{+}{=}
    \partial_2 \phi_k ( v_{a[k]} , v_{b[k]} ) \bar{v}_k
\) \\
\end{tabbing}
Differentiating the algorithm above with respect to $x$,
and using the forward mode equation for
$\dot{v}_k$ in Eq~\ref{EquationForwardHessian},
we obtain
\begin{tabbing}
\hspace{2em}\=\hspace{2em}\=\hspace{2em}\=\hspace{4em}\= \kill
\> \B{for} $k = 1 , \ldots , \ell$ \B{set} $\dot{\bar{v}}_k = 0$ \\
\> \B{for} $k = \ell , \ldots , n+1$ \\
\> \> \B{comment} differentiate setting of $\bar{v}_{a[k]}$
\\
\> \> \B{set} \(
\dot{\bar{v}}_{a[k]} \stackrel{+}{=}
    \partial_1 \phi_k ( v_{a[k]} , v_{b[k]} ) \dot{\bar{v}}_k
\) \\
\> \> \B{set} \(
\dot{\bar{v}}_{a[k]} \stackrel{+}{=}
    \partial_{1,1} \phi_k ( v_{a[k]} , v_{b[k]} ) \bar{v}_k \dot{v}_{a[k]}
\) \\
\> \> \B{set} \(
\dot{\bar{v}}_{a[k]} \stackrel{+}{=}
    \partial_{1,2} \phi_k ( v_{a[k]} , v_{b[k]} ) \bar{v}_k \dot{v}_{b[k]}
\) \\
\> \> \B{comment} differentiate setting of $\bar{v}_{b[k]}$
\\
\> \> \B{set} \(
\dot{\bar{v}}_{b[k]} \stackrel{+}{=}
    \partial_2 \phi_k ( v_{a[k]} , v_{b[k]} ) \dot{\bar{v}}_k
\) \\
\> \> \B{set} \(
\dot{\bar{v}}_{b[k]} \stackrel{+}{=}
    \partial_{1,2} \phi_k ( v_{a[k]} , v_{b[k]} ) \bar{v}_k \dot{v}_{a[k]}
\) \\
\> \> \B{set} \(
\dot{\bar{v}}_{b[k]} \stackrel{+}{=}
    \partial_{2,2} \phi_k ( v_{a[k]} , v_{b[k]} ) \bar{v}_k \dot{v}_{b[k]}
\) \\
\end{tabbing}
Suppose that in Eq~\ref{EquationForwardHessian}
$\dot{x}$ is the $j$-th elementary vector.
It follows that $\dot{\bar{x}}_p = \partial_{p,j} g(x)$
for $p = 1 , \ldots n$.
We claim that at the beginning of the iteration $k$,
in the algorithm above and in Algorithm~\ref{AlgorithmReverseHessianSparsity},
for $p = 1 , \ldots , k$
\[
    \dot{\bar{v}}_p \neq 0 \Rightarrow j \in M_p \; .
\]
Proving this claim will complement the proof of the theorem.
For the first iteration, $k = \ell$ and $\dot{\bar{v}}_p = 0$ for all $p$.
Hence the claim is true for $k = \ell$.
Suppose the claim is true at the beginning of the $k$-th iteration,
it suffices to show it is true at the beginning of iteration $k-1$.
If $p \neq a[k]$ and $p \neq b[k]$
then $\dot{\bar{v}}_p$ and $M_p$ are the same at the beginning of
iteration $k$ and $k-1$, so we are done.
The two cases $p = a[k]$ and $p = b[k]$ are symmetric.
It suffices to show the case $p = a[k]$; i.e.,
at the end of iteration $k$
\[
    \dot{\bar{v}}_{a[k]} \neq 0 \Rightarrow j \in M_{a[k]} \; .
\]
If $\dot{\bar{v}}_{a[k]} \neq 0$ at the beginning of iteration $k$
then by induction $ j \in M_{a[k]} $ at the beginning of iteration $k$
and by Algorithm~\ref{AlgorithmReverseHessianSparsity} it also
true at the end of iteration $k-1$.

Consider the remaining case where
$\dot{\bar{v}}_{a[k]} = 0$ at the beginning of iteration $k$ and
$\dot{\bar{v}}_{a[k]} \neq 0$ at the end of iteration $k$.
This implies that the right hand side was non-zero in one of the
three assignments to $\dot{\bar{v}}_{a[k]}$ above.
This in turn implies that $d_k$ is true
(otherwise $\bar{v}_k$ and $\dot{\bar{v}}_k$ would be zero).
Suppose the first assignment to $\dot{\bar{v}}_{a[k]}$ is non-zero,
\[
0 \neq \partial_1 \phi_k ( v_{a[k]} , v_{b[k]} ) \dot{\bar{v}}_k
\; .
\]
This implies that $\dot{\bar{v}}_k \neq 0$ which, by induction, implies
that $j \in M_k$ which, by Algorithm~\ref{AlgorithmReverseHessianSparsity},
implies $j \in M_{a[k]}$ at the end of iteration $k$.
This completes the case where the first assignment to
$\dot{\bar{v}}_{a[k]}$ is non-zero.

Suppose the second assignment to $\dot{\bar{v}}_{a[k]}$ is non-zero,
\[
0 \neq \partial_{1,1} \phi_k ( v_{a[k]} , v_{b[k]} )  \bar{v}_k \dot{v}_{a[k]}
\; .
\]
This implies that all three terms in the product on the
right hand side are non-zero.
Hence $\dot{v}_{a[k]} = \partial_j v_{a[k]} (x)$
is non-zero and $j \in X_{a[k]}$.
Furthermore $\phi_k$ is left nonlinear.
Hence, at the end of iteration $k$,
$X_{a[k]} \subset M_{a[k]}$ and $j \in M_{a[k]}$
This completes the case where the second assignment to
$\dot{\bar{v}}_{a[k]}$ is non-zero.

Suppose the third assignment to $\dot{\bar{v}}_{a[k]}$ is non-zero,
\[
0 \neq \partial_{1,2} \phi_k ( v_{a[k]} , v_{b[k]} )  \bar{v}_k \dot{v}_{b[k]}
\; .
\]
This implies that all three terms in the product on the
right hand side are non-zero.
Hence $\dot{v}_{b[k]} = \partial_j v_{b[k]} (x)$
is non-zero and $j \in X_{b[k]}$.
Hence, at the end of iteration $k$,
$X_{b[k]} \subset M_{a[k]}$ and $j \in M_{a[k]}$.
This completes the case where the third assignment to
$\dot{\bar{v}}_{a[k]}$ is non-zero.
Q.E.D.

\begin{remark}
\label{RemarkEdgePushing}
The edge pushing algorithm
\cite[Algorithm 3.1]{Gower2014} is an alternative reverse mode algorithm
for computing Hessian sparsity patterns.
It is fundamentally different from the reverse mode Hessian sparsity
algorithm in this paper because it does not use first order forward mode
sparsity patterns to reduce the computed index set sizes to the number of
independent variables that are of interest.
\end{remark}

\subsection{Subgraph Sparsity}
\label{SubgraphSparsity}
We are given a function $f : \B{R}^n \rightarrow \B{R}^m$ and define
$g : \B{R}^n \rightarrow \B{R}$ by
\[
g(x) = \sum_{i=1}^m w_i f_i (x) \; .
\]
Using reverse mode we can compute $g^{(1)} (x)$ in $O( \ell )$ operations.
In addition we can obtain the corresponding computational graph; for example
see \cite[Tape T2 on pp. 7]{Kristensen2016}, \cite[Section 3.1]{Wang2016}.
We use this computational graph to define the function
$h : \B{R}^n \rightarrow \B{R}^n$
\[
h(x) = g^{(1)} (x) = \sum_{i=1}^m w_i f_i^{(1)} (x) \; .
\]
The Jacobian of $h(x)$ is the Hessian of $g(x)$.
We can apply the reverse subgraph Jacobian algorithm to
the computational graph for $h(x)$ to obtain the
sparsity pattern for the Hessian $g^{(2)} (x)$.

\subsection{Computing Hessians}
Once a sparsity pattern for a Hessian is available,
the values in the Hessian are computed in a manner similar to how the
Jacobians are calculated; see Section~\ref{SectionComputingJacobians}.

\section{Experiments}
\label{SectionExperiments}
The speed tests reported below were run using the following hardware
and software:

\begin{tabular}{rl}
processor:                                     & i5-3470 3.2GHz 64bit \\
operating system:                              & fedora 34 \\
Linux kernel:                                  & 5.13.6 \\
compiler:                                      & clang 12.0.0 \\
memory:                                        & 8GB \\
disk:                                          & ST500DM002-1BD14 \\
{\small ADOL-C}       \cite{Griewank1999}:     & git hash 25a69c4 \\
{\small CppAD}        \cite{BellCppAD}:        & git hash 5c86b18 \\
{\small ColPack}      \cite{Gebremedhin2013}:  & version v1.0.10 \\
\end{tabular}

The {\small ADOL-C} git hash corresponds to its master branch on 2020-03-31,
the {\small CppAD} git hash corresponds to its master branch on 2020-05-23.

\subsection{CSV File}
Each run of the speed test program adds a row to a csv file with
the following columns:

\medskip
\textit{KB}:
This is the average memory usage, in kilobytes,
where each kilobyte is 1000 bytes (not 1024 bytes).
A separate run of the speed test program by
valgrind's massif tool is used to determine these values.

\medskip
\textit{implement}:
\begin{itemize}
\setlength{\itemindent}{2.0em}
\item[\texttt{adolc}]
    A forward or reverse algorithm implemented by the {\small ADOL-C} package
\item[\texttt{cppad}]
    A forward or reverse algorithm implemented by the {\small CppAD} package
\item[\texttt{subgraph}]
    A {\small CppAD} implementation of
    Algorithm \ref{AlgorithmReverseSubgraphJacobianSparsity}.
    Note that the potentially faster Algorithm
    \ref{AlgorithmReverseSubgraphJacobianSubgraph} is not implemented in
    {\small CppAD} and therefore not part of the test.
\end{itemize}

\medskip
\textit{problem}:
This is the \texttt{minpack2} test source code used to compute $f(x)$;
see \cite{Averick1992}.
The source code was converted to C using \texttt{f2c} and then
to {\small C++} so it could be used with {\small ADOL-C} and {\small CppAD}.
The available \textit{problem} values are:
\begin{itemize}
\setlength{\itemindent}{2.0em}
\item[\texttt{dficfj}]  The flow in a channel Jacobian problem
\item[\texttt{dierfj}]  The incompressible elastic rod Jacobian problem
\item[\texttt{deptfg}]  The elastic-plastic torsion Hessian problem
\item[\texttt{dgl1fg}]  One Dimensional Ginzburg-Landau Hessian problem
\end{itemize}

\medskip
\textit{colpack}:
If true,
the \texttt{ColPack} package
was used to solve the coloring sub-problem.
Otherwise a greedy distance two coloring algorithm
(inspired by the algorithm \cite[Algorithm 3.2]{Gebremedhin2005} and
implemented in the {\small CppAD} package)
is used.
In either case, if this is a Hessian problem,
a special version of the coloring algorithm
that takes advantage of the symmetry, is used.
The \textit{colpack} option must be true
when \textit{implement} is \texttt{adolc}.
It must be false when \textit{implement} is \texttt{subgraph}
because it does not use coloring.

\medskip
\textit{indirect}:
If this is true, an indirect method was used to get more compression
out of the coloring problem.
It can only be true when \textit{implement} is \texttt{adolc}
and \textit{problem} corresponds to computing a Hessian.
This must be false when \textit{implement} is \texttt{cppad}
because it does not support this recovery option.
It must also be false when \textit{implement} is \texttt{subgraph}
because it does not use a compression technique.

\medskip
\textit{optimize}:
Was the computation graph was optimized.
This must be false when \textit{implement} is \texttt{adolc}
because it does not have this option.

\medskip
\textit{setup}:
Does the \textit{sec} result include the setup time; e.g.,
recording the graph, computing sparsity, coloring, compression, and
optimization of the graph.
If the computational graph (see Section~\ref{SectionComputationalGraph})
corresponding to a function does not depend on the argument to the function,
the setup operations can be preformed once and re-used for any argument value.
Note that optimization is not included in the setup time when
\textit{implement} is \texttt{adolc}.
Coloring and compression are not included when
\textit{implement} is \texttt{subgraph}.

\medskip
\textit{reverse}:
If true (false) reverse mode (forward mode)
was used for computing both the sparsity pattern and derivative values.
This option must be true when
\textit{implement} is \texttt{subgraph}.

\medskip
\textit{onepass}:
If true, the derivative values were computed using
one pass of graph with multiple directions at the same time.
Otherwise, each directional derivative is computed using a separate pass.
This option must be false when
\textit{implement} is \texttt{subgraph}.
The \textit{onepass} option must be true when
\textit{implement} is \texttt{adolc} and \textit{reverse} is false.

\medskip
\textit{n}:
This is the size of the domain space for $f(x)$.

\medskip
\textit{m}:
This is the size of the range space for $f(x)$.
If $m = 1$ ($m > 1$), this is a Hessian (Jacobian) problem.

\medskip
\textit{nnz}:
This is the number of possibly non-zero values in sparse matrix calculated
by the implementation.
For a Jacobian problem, the matrix is the entire Jacobian.
For a Hessian problem, the matrix is the upper triangle of the Hessian.
Note that one implementation may compute a more efficient sparsity pattern
than another (have fewer possibly non-zero values).

\medskip
\textit{sec}:
This is the number of seconds for each calculation
of the sparse Jacobian or Hessian.

\subsection{Result Tables}
There is one table for each \textit{problem} and value for \textit{setup}.
For each \textit{implement}, only the combination of options that result in the
smallest \textit{sec} (smallest time and fastest execution)
is included in the table.
Given this selection,
options that are the same for all the implementations in a table
are reported in the caption at the top of the table.
The other options are reported in each row of the table.
The rows for each table are sorted
so that the \textit{sec} column is monotone non-decreasing.

\subsubsection{Jacobian Without Setup}
\begin{table}[tbhp]
{ \footnotesize
\caption{\textit{problem}=\texttt{dficfj},
\textit{setup}=\texttt{false},
\textit{indirect}=\texttt{false},
\textit{n}=\texttt{3200},
\textit{m}=\texttt{3200},
\textit{nnz}=\texttt{24787}
}
\label{TableDficfjNoSetup}
\begin{center}
\begin{tabular}{rrrrrrl}
KB & implement & colpack & optimize & reverse & onepass & sec\\
12108 & adolc & true & false & false & true & 0.00186 \\
13361 & cppad & true & true & false & true & 0.00233 \\
6480 & subgraph & false & true & true & false & 0.00586
\end{tabular}
\end{center}
}

\bigskip
{ \footnotesize
\caption{\textit{problem}=\texttt{dierfj},
\textit{setup}=\texttt{false},
\textit{indirect}=\texttt{false},
\textit{n}=\texttt{3003},
\textit{m}=\texttt{3003},
\textit{nnz}=\texttt{31600}
}
\label{TableDierfjNoSetup}
\begin{center}
\begin{tabular}{rrrrrrl}
KB & implement & colpack & optimize & reverse & onepass & sec\\
11561 & adolc & true & false & false & true & 0.0015 \\
5797 & cppad & true & true & false & true & 0.00182 \\
4049 & subgraph & false & true & true & false & 0.00411
\end{tabular}
\end{center}
}

\end{table}
Table~\ref{TableDficfjNoSetup} (Table~\ref{TableDierfjNoSetup})
compares the time, \textit{sec}, and memory, \textit{KB},
required to compute the Jacobian where \textit{problem} is
\texttt{dficfj} (\texttt{dierfj}) and \textit{setup} is false.
The \texttt{adolc} and \texttt{cppad} implementations
use the same algorithms, hence their similar run times is not surprising.
The \texttt{subgraph} implementation takes about twice as long.
(The \texttt{subgraph} implementation does not require a sparsity calculation
or solving a coloring sub-problem and these tasks are part of the setup
for the \texttt{adolc} and \texttt{cppad} implementations.)
The number of possibly non-zeros in the Jacobian sparsity pattern, \textit{nnz},
depends on the problem but does not depend on the implementation; i.e.,
the sparsity pattern efficiency is the same for all the implementations.
The fastest option choices for the \texttt{cppad} implementation are
\textit{optimize} true,
\textit{reverse} false, and
\textit{onepass} true.
Using \textit{colpack} true with the \texttt{cppad} implementation is faster
for one problem and slower for the other.
The \textit{indirect} true option is not available for Jacobian problems.
The \textit{onepass} true option results in more memory usage
but we have not included those details in the result tables.
The \texttt{subgraph} implementation uses less memory
than the other implementations.

\subsubsection{Jacobian With Setup}
\begin{table}[tbhp]
{ \footnotesize
\caption{\textit{problem}=\texttt{dficfj},
\textit{setup}=\texttt{true},
\textit{indirect}=\texttt{false},
\textit{optimize}=\texttt{false},
\textit{n}=\texttt{3200},
\textit{m}=\texttt{3200},
\textit{nnz}=\texttt{24787}
}
\label{TableDficfjYesSetup}
\begin{center}
\begin{tabular}{rrrrrl}
KB & implement & colpack & reverse & onepass & sec\\
9807 & adolc & true & false & true & 0.0109 \\
18430 & cppad & true & false & true & 0.013 \\
8859 & subgraph & false & true & false & 0.0141
\end{tabular}
\end{center}
}

\bigskip
{ \footnotesize
\caption{\textit{problem}=\texttt{dierfj},
\textit{setup}=\texttt{true},
\textit{indirect}=\texttt{false},
\textit{optimize}=\texttt{false},
\textit{n}=\texttt{3003},
\textit{m}=\texttt{3003},
\textit{nnz}=\texttt{31600}
}
\label{TableDierfjYesSetup}
\begin{center}
\begin{tabular}{rrrrrl}
KB & implement & colpack & reverse & onepass & sec\\
9585 & adolc & true & false & true & 0.00862 \\
5151 & subgraph & false & true & false & 0.00887 \\
13645 & cppad & false & false & true & 0.0116
\end{tabular}
\end{center}
}

\end{table}
Table~\ref{TableDficfjYesSetup} (Table~\ref{TableDierfjYesSetup})
is similar to
Table~\ref{TableDficfjNoSetup} (Table~\ref{TableDierfjNoSetup})
with the difference being that
\textit{setup} is true instead of false.
The time for the \texttt{adolc}, \texttt{cppad} and \texttt{subgraph}
implementations are all close.
The fastest option choices for the \texttt{cppad} implementation are
\textit{optimize} false,
\textit{reverse} false, and
\textit{onepass} true.
Using \textit{colpack} true with the \texttt{cppad} implementation is faster
for one problem and slower for the other.
The \texttt{subgraph} implementation uses less memory
than the other implementations.

\subsubsection{Hessian Without Setup}
\begin{table}[tbhp]
{ \footnotesize
\caption{\textit{problem}=\texttt{deptfg},
\textit{setup}=\texttt{false},
\textit{onepass}=\texttt{false},
\textit{n}=\texttt{3600},
\textit{m}=\texttt{1}
}
\label{TableDeptfgNoSetup}
\begin{center}
\begin{tabular}{rrrrrrrl}
KB & implement & colpack & indirect & optimize & reverse & nnz & sec\\
5207 & subgraph & false & false & true & true & 10680 & 0.00499 \\
4364 & cppad & true & false & true & false & 10680 & 0.00523 \\
15115 & adolc & true & true & false & false & 14161 & 0.0226
\end{tabular}
\end{center}
}

\bigskip
{ \footnotesize
\caption{\textit{problem}=\texttt{dgl1fg},
\textit{setup}=\texttt{false},
\textit{onepass}=\texttt{false},
\textit{n}=\texttt{5000},
\textit{m}=\texttt{1},
\textit{nnz}=\texttt{10000}
}
\label{TableDgl1fgNoSetup}
\begin{center}
\begin{tabular}{rrrrrrl}
KB & implement & colpack & indirect & optimize & reverse & sec\\
6086 & cppad & true & false & true & true & 0.00535 \\
9806 & subgraph & false & false & true & true & 0.00982 \\
16679 & adolc & true & true & false & false & 0.0243
\end{tabular}
\end{center}
}

\end{table}
Table~\ref{TableDeptfgNoSetup} (Table~\ref{TableDgl1fgNoSetup})
compares the time, \textit{sec}, and and memory, \textit{KB},
required to compute the Hessian where \textit{problem} is
\texttt{deptfg} (\texttt{dgl1fg})
and \textit{setup} is false.
The \texttt{cppad} and \texttt{subgraph} implementations have similar times
and the \texttt{adolc} implementation takes twice as long.
The number of possibly non-zeros in the \texttt{adolc} implementation
for the \texttt{deptfg} problem is significantly
larger than in the other implementations; i.e.,
its sparsity pattern is not as efficient as the other implementations.
The fastest option choices for the \texttt{cppad} implementation are
\textit{colpack} true,
\textit{optimize} true, and
\textit{reverse} true.
The \textit{indirect} true option is not available
with the \texttt{cppad} implementation.
No implementation uses less memory for both problems.

\subsubsection{Hessian With Setup}
\begin{table}[tbhp]
{ \footnotesize
\caption{\textit{problem}=\texttt{deptfg},
\textit{setup}=\texttt{true},
\textit{onepass}=\texttt{false},
\textit{n}=\texttt{3600},
\textit{m}=\texttt{1}
}
\label{TableDeptfgYesSetup}
\begin{center}
\begin{tabular}{rrrrrrrl}
KB & implement & colpack & indirect & optimize & reverse & nnz & sec\\
7809 & subgraph & false & false & true & true & 10680 & 0.0258 \\
15299 & adolc & true & true & false & false & 14161 & 0.0453 \\
4921 & cppad & true & false & true & false & 10680 & 0.0843
\end{tabular}
\end{center}
}

\bigskip
{ \footnotesize
\caption{\textit{problem}=\texttt{dgl1fg},
\textit{setup}=\texttt{true},
\textit{onepass}=\texttt{false},
\textit{n}=\texttt{5000},
\textit{m}=\texttt{1},
\textit{nnz}=\texttt{10000}
}
\label{TableDgl1fgYesSetup}
\begin{center}
\begin{tabular}{rrrrrrl}
KB & implement & colpack & indirect & optimize & reverse & sec\\
33709 & subgraph & false & false & false & true & 0.0603 \\
7917 & cppad & true & false & true & false & 0.111 \\
19144 & adolc & true & true & false & false & 0.114
\end{tabular}
\end{center}
}

\end{table}
Table~\ref{TableDeptfgYesSetup} (Table~\ref{TableDgl1fgYesSetup})
is similar to
Table~\ref{TableDeptfgNoSetup} (Table~\ref{TableDgl1fgNoSetup})
with the difference being that
\textit{setup} is true instead of false.
The \texttt{subgraph} implementation is the fastest for this case
and the \textit{cppad} implementation is significantly slower.
The fastest option choices for the \texttt{cppad} implementation are
\textit{colpack} true,
\textit{optimize} false, and
\textit{reverse} true.
The \texttt{adolc} implementation uses less memory
than the other implementations.

\subsubsection{Reproducing Results}
The source code that produced the results corresponds to
the tag \texttt{20210803} of the following git repository:
\begin{center}
{ \small
\url{https://github.com/bradbell/sparse_ad}
}.
\end{center}

\section{Conclusions}
If the computational graph (see Section~\ref{SectionComputationalGraph})
corresponding to a function does not depend on the argument to the function,
the setup can be done once and reused many times.
In the other case,
when the setup operations are computed for each argument to a sparse Jacobian,
all the implementation have similar timing results.
When computing the setup operations for each argument to a sparse Hessian,
the \texttt{subgraph} implementation is significantly faster and
\texttt{cppad} is significantly slower.

Further testing is called for.
We have provided a GitHub repository
with the source code used to obtain the results in this paper.
This facilitates reproduction of the results as well as
extension of the tests to other cases.
Other cases include different
computer hardware,
operating systems,
compilers, and
versions of the AD packages.
The tests can also be extended to other AD packages
as well as other problems and problem sizes.
The tests also provide example source code that implements the
algorithms presented in this paper.

\bibliography{sparse}

\end{document}